\documentclass[%
 reprint,
nofootinbib,
 amsmath,amssymb,
 aps,
]{revtex4-2}

\usepackage{hyperref}
\usepackage{url}
\usepackage{graphicx}
\usepackage{dcolumn}
\usepackage{bm}

\begin{document}


\title{IPPOG: a global network for particle physics outreach and education}



\collaboration{IPPOG Collaboration}



\date{\today}

\begin{abstract}
We present the International Particle Physics Outreach Group (IPPOG), a global network dedicated to connecting students, educators, and the general public with the world of particle physics.  In this paper, we outline the need to bridge the existing gap between the particle physics community and the wider audience, and we present the solutions that IPPOG has implemented to overcome it through three pillar Activities: the International Masterclasses and the Global Cosmics hands-on activities network, which have engaged together over 200\,000 high-school students to date, and the curation of an Outreach Resource Database and web portal.

\end{abstract}

\maketitle


\section{Why Outreach Matters: Bridging Particle Physics and the Public}
\label{sec:why-outreach}

Modern physics, and in particular particle physics, permeate the world. The usage of quantum mechanical phenomena surrounds us in ways that largely go unnoticed. To understand the Universe that surrounds us physics is quintessential. And yet, there is a gap between the research world and the public-at-large. In many countries it is fairly common to receive the first exposure to quantum mechanics only towards the end of bachelor degrees in physics, and curricula in high schools rarely go beyond classical mechanics and electromagnetism.

Outreach initiatives are therefore indispensable to bridge the divide between cutting-edge research and the classroom. The International Particle Physics Outreach Group (IPPOG) provides educational and communication opportunities such as the International Masterclasses \cite{ippog-imc}, virtual laboratory tours, and hands-on cosmic ray data collection and analysis \cite{GlobalCosmics} to bring real research and the enthusiasm of the researchers into teaching environments. IPPOG serves then as the front line of the particle physics world with society, connecting a network of physicists, teachers, and communication experts with interested audiences, with the aim of demystifying complex concepts, illustrating the societal applications of particle physics, and nurturing scientific curiosity and critical thinking \cite{ResourcesPortal}.

In addition, outreach is of paramount importance for inclusion and diversity, and thus providing access to STEM opportunities to different countries worldwide. The effort of IPPOG is geared to achieve the possibility for learners from diverse and geographically different backgrounds to engage with particle physics. This effort is crucial to foster the physics community of tomorrow, and to lead to the creation of a scientifically informed public.

The purpose of this article is to introduce IPPOG and its activities, in order to provide a point of contact for teachers and the wider public.
 
\section{IPPOG’s History \& Mission}
\label{sec:history}

The International Particle Physics Outreach Group (IPPOG) is a global network of scientists, educators, and communication specialists dedicated to informal science education and public engagement in particle physics. It is the successor of the early European Particle Physics Outreach Group (EPPOG), which was formed in 1997 with the support of both the European Committee for Future Accelerators (ECFA) and the High-Energy and Particle Physics Board of the European Physical Society (EPS)
. EPPOG was created with the purpose of fostering outreach expertise, pooling resources, and carrying out communication activities for particle physics to schools and the public. In 2005, EPPOG was expanded  by launching the International Particle Physics Masterclass program, allowing students worldwide to analyse real experimental data, and thus familiarise themselves with activities that mostly pertained to the research environment until then \cite{ippog-imc2005}. EPPOG was officially renamed IPPOG in 2011, to reflect its growing international spread. Finally, it became a formal collaboration in 2016, reaching its current maturity stage.

\begin{figure}[h]
\includegraphics[width=0.95\columnwidth]{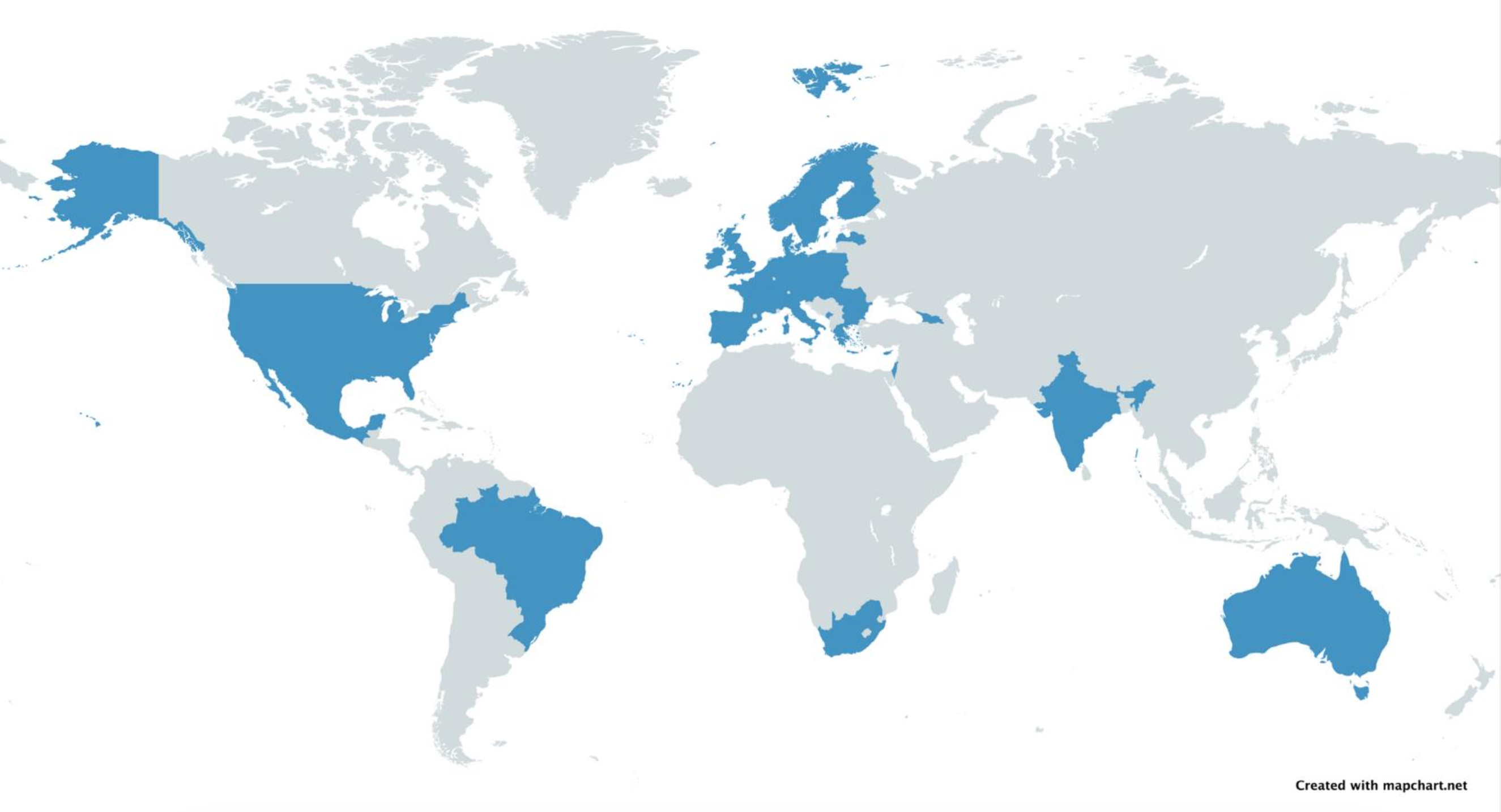}%
\caption{\label{fig:map}IPPOG Collaboration map.}
\end{figure}

IPPOG already consisted of 42 members at the end of 2024: 34 countries, 7 experiments and an international laboratory (CERN). Two national laboratories (DESY and GSI) are associate members. In addition, partnerships have been  developed with other outreach groups such as QuarkNet \cite{QuarkNet} in the United States. A map of which countries are members of IPPOG is shown in Fig.~\ref{fig:map}. 

\begin{figure}[h]
\includegraphics[width=0.95\columnwidth]{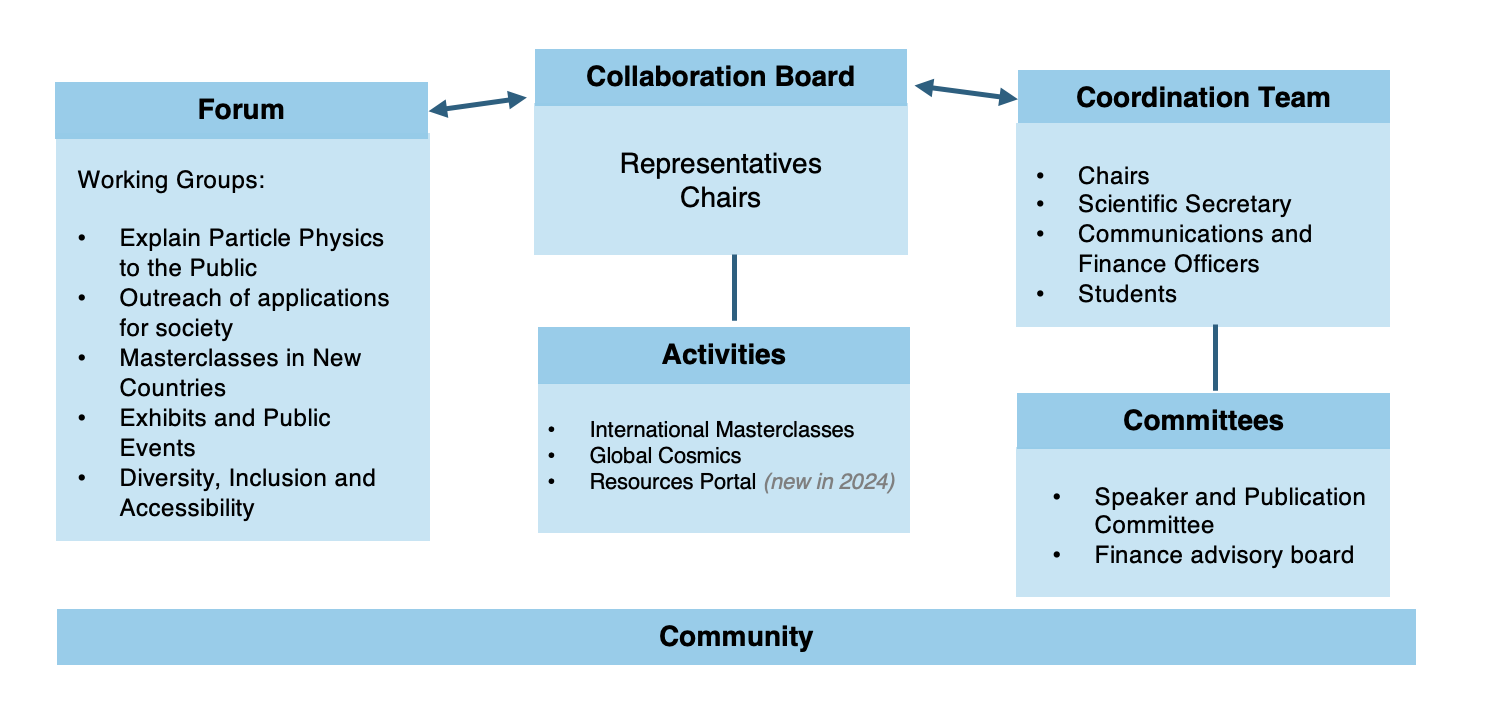}%
\caption{\label{fig:structure} IPPOG Collaboration structure.}
\end{figure}

The structure of the IPPOG Collaboration is presented in Fig.~\ref{fig:structure}. The Collaboration Board, which includes one representative from each member and associate member organizations, meets twice a year to discuss and vote on IPPOG-related matters, including the election of the Chairs and of the Activities Coordinators. The Coordination Team is the executive body of the Collaboration. It is run by the Chairs with a dedicated team of officers and students, and two specialized committees (speakers and publication committee, finances advisory board). The Forum is the IPPOG discussion body, whose members are nominated by the representatives: at the end of 2024, it counted 183 individuals. Informal working groups foster the development of specific resources, strategies and best practices.  Finally, presentations at conferences, shared projects, and informal exchanges connect IPPOG with the wider high-energy physics outreach community. 
The mission of IPPOG is twofold: its structure and extended partnerships allow it to organize international events and programs, and its deep roots shed light on the local diversity and engagement of research institutes. The IPPOG Forum gathers contributions from experiments from all over the world, in the broader meaning of particle physics, from CERN (Switzerland) to Fermilab (USA) and KEK (Japan), from DESY (Germany) to the Pierre Auger Observatory (Argentina) and HAWC (Mexico). This allows access to open data sets, facilities, and expertise belonging to different branches of physics, from particle to high-energy nuclear physics and astroparticle physics. 

Twice a year, members of the community are invited to present their activities to the Collaboration in \textit{Success Stories} sessions. Contributions are grouped into an annual report, which is an evolving snapshot of the state-of-the-art \cite{IPPOGgeneralreport, IPPOG-membersreport}. 
\section{Three international Activities}
\label{sec:activities}

IPPOG carries out its international program through three different pillar activities, each designed to interact with students, educators, and the public through different strategies. 
\subsection{International Masterclasses}
IPPOG has been inviting high school students to become \textit{scientists for a day} through the International Masterclasses program since 2005 \cite{Johansson:2006,Kobel:2007}. The participants have the opportunity to analyse real experimental data and to present their results in a videoconference format. 

The current scientific portfolio \cite{ippog-imc} grew in steps, starting with measurements from LEP experiments \cite{LEP} and followed in 2012 by the four large LHC experiments: ATLAS (Z and W boson-related measurements), CMS (Z and W boson-related measurements), ALICE (quark-gluon plasma-related observables), and LHCb (heavy-flavor measurements) \cite{ippog-imc2014, ATLASMasterclasses, CMSMasterclasses, ALICEMasterclasses, LHCbD0LifetimeMasterclass}. Over the years the Masterclass program became more and more encompassing, going beyond the original CERN-related theme to embrace experiments in different continents: Belle~II at KEK (flavour physics measurements) \cite{BELLEIIMasterclasses}, neutrino physics at Fermilab with MINER$\,\nu$A \cite{MINERVAMasterclass} and NO$\nu$A \cite{NOVAMasterclass2024}. The Pierre Auger Observatory enriched the landscape with cosmic ray sessions \cite{AugerMasterclasses, AugerMasterclassesWeb}, while cosmic neutrino and gravitational wave exercises are under development. A medical physics particle therapy masterclass \cite{ParticleTherapyMasterClass2020} connects fundamental research with its application.

The scale of the annual international campaign has grown remarkably from approximately 3\,000 students in 18 countries in 2005 to over 13\,000 students across 60 countries at 225 host institutions in 2023, demonstrating the high demand for this type of activities and the excitement of the students to experience particle physics and the research world. In 2024 alone, 14\,700 participants attended, at the end of a day of visits and activities, the 110 online discussion sessions organised by moderators at CERN, Fermilab, GSI, KEK, and the Pierre Auger site,  \cite{EngagingWorld,IPPOGgeneralreport}. The coordination effort required to maintain such a large scale program requires the support from all experiments and hundreds of volunteers on a rotation system.
\subsection{Global Cosmics}
The Global Cosmic group invites educators and students to embark on hands‑on sessions using setups suitable to be deployed as tabletop designs. Projects have multiplied other the course of the last decade, with the IPPOG group serving as an exchange hub. 

In some of them the participants receive step-by-step instructions to build and assemble cosmic ray detectors, often using scintillator materials and photomultipliers. In others, open source software is prepared to facilitate data acquisition and data to analyse, as for example in the recent Cosmic Piano project, that introduces both a hardware and data analysis experience \cite{TejedaMunoz2025CosmicPiano}. 
Key examples are, in Germany, the CosMO and Kamiokannen experiments, designed as part of the nationwide Netzwerk Teilchenwelt, as well as a platform called Cosmic@Web where collected data can be analyzed \cite{CosMO,Cosmic@Web}. The Japanese Accel Kitchen initiative has distributed compact cosmic-ray detectors to more than 200 high school students, primarily across Asia, enabling them to assemble and operate the devices at home with the support of a network of undergraduate and graduate students \cite{Accel}. In the US, Mexico and Japan, teams were involved in muon tomography projects, e.g. to search for hidden chambers at the great pyramid at Chichen Itza (Mexico) or to study ancient burial mounds (Japan).

\begin{figure}[h]
\includegraphics[width=0.95\columnwidth]{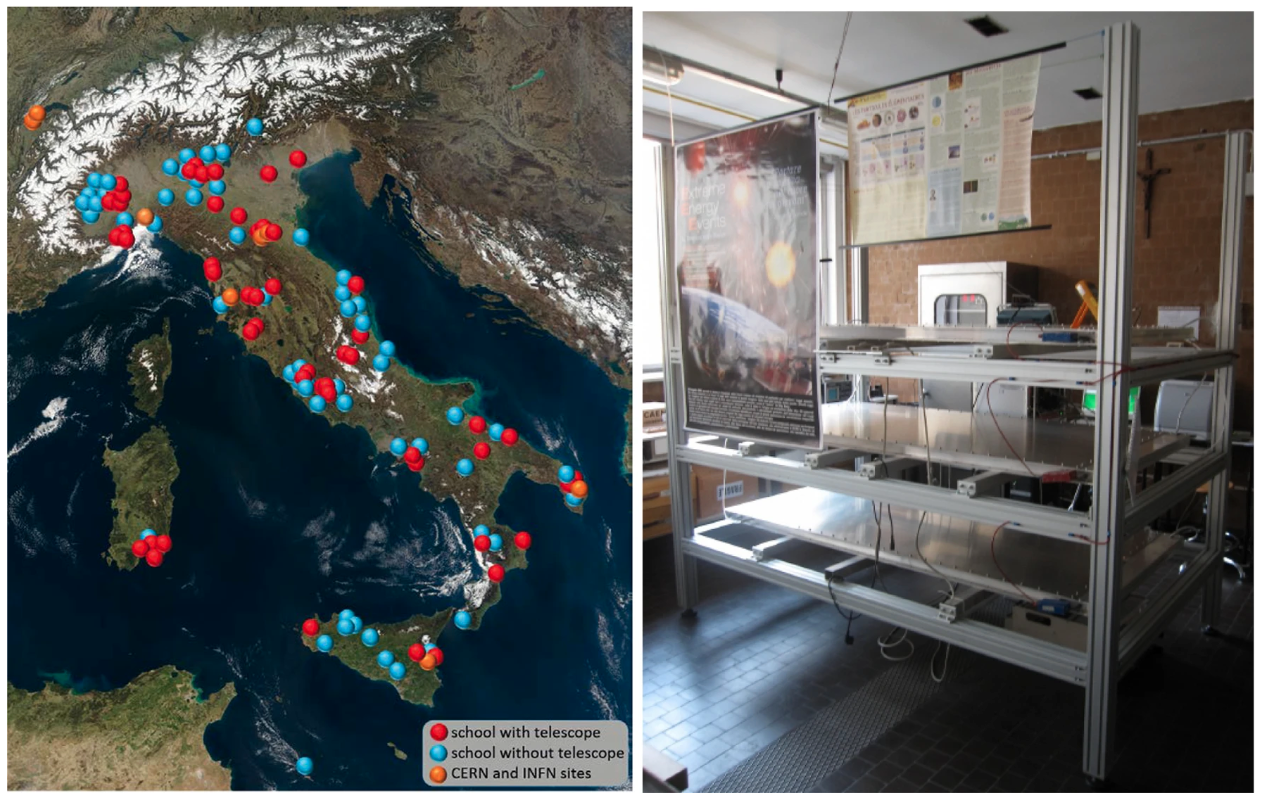}%
\caption{\label{fig:eee}The EEE network (left). Red and orange dots indicate the locations of the EEE stations in high schools and at CERN or INFN laboratories, respectively. Cyan dots mark schools participating in the project without a telescope. On the right, one such telescope is shown. Figure taken from \cite{Pisano:2022}. }
\end{figure}

Long term collaborations are organised with schools. French institutes have developed \textit{Cosmodetecteurs} and  \textit{COSMIX suitcases} shared with schools, training programs, and even installed a full set of detectors on top of the Pic du Midi Observatory \cite{cosmixcase}. In the US, QuarkNet \cite{QuarkNet} organizes multiple cosmic ray activities through long-term collaborations between high school teachers and scientists. During the International Muon Week \cite{MuonWeek}, paired schools collect and analyse data and share the results they have obtained. In Italy, the INFN Outreach Cosmic Ray Activities (OCRA) \cite{Aramo:2020les} project involves 24 INFN divisions, offering activities for both students and teachers.  The Extreme Energy Events (EEE), shown in Fig.~\ref{fig:eee}, is one of the most successful projects. It is a large area array based on multigap resistive plate chambers (MRPCs), where students are directly involved in the assembly, maintenance and data analysis.
In the same spirit as the masterclasses, the International Cosmic Day (ICD) \cite{ICD}, organized each year by DESY, is the occasion to engage a larger audience. High school students from across the world have the opportunity to carry out hands-on measurements of the cosmic ray flux and to work with real astroparticle physics data, while scientists have the chance to share their experiences. Each participating institute chooses the format and content of the day, while videoconferences with groups in other countries, drawings, and photographic competitions add a distinctive flavour that enriches the experience.

\vspace{1em}
\subsection{Offering Outreach Resources}
Since the creation of IPPOG the reach and variety of outreach projects have grown significantly, as have the visibility and recognition of the importance of public engagement. The material that IPPOG pioneers developed and shared in the early days of the network is still available in a solid Resource Database (RDB) \cite{RDB} curated for its 25th anniversary. 

With the growth of websites and social media, most of the experiments now offer, via websites, a specific set of links, tools and material in multiple languages \cite{WebSites}. Every high-energy physics conference also now proposes well-attended and lively outreach parallel sessions and plenary talks. Offering an accessible and structured entry point for all the resources is not an easy task, yet it is very important for newcomers in search for inspiration and contacts. Teachers who are invited and trained in large laboratories, such as CERN \cite{TeachersProgramme}, also show their interest in follow-up activities and resources. The last ingredient missing was thus the presentation to the public of the specific projects. Since 2019, IPPOG has hosted biannual \textit{success stories} sessions that showcase highlights from our global outreach community. In 2025 these presentations were consolidated on a dedicated web portal (see \cite{ResourcesPortal}), where authors tag their own submissions across topic, format, audience, and language. This searchable repository not only disseminates ideas and best practices, but also recognizes the efforts of our colleagues worldwide.


\section{\label{sec
}Conclusions}

From the network of the early days to a structured collaboration, IPPOG has, over the course of nearly three decades, established a solid framework that connects the academic and research community to learners, educators, and the general public. It has engaged over 200\,000 students and teachers through its Three Activities, i.e. the International Masterclasses, the Global Cosmics initiative, and the Resource Database, fostering not only educators but also a global community of citizen scientists. By deepening the dialogue between particle physicists and society, IPPOG aims to nurture future generations of scientists, and to ensure public support for the fundamental research in particle physics.


\begin{acknowledgments}
The IPPOG Collaboration gratefully acknowledges the invaluable contribution, skills and dedication of the colleagues, students and volunteers without whom particle physics outreach would not exist. It is supported by 37 institutions around the world, via financial contributions, personnel or support.

\end{acknowledgments}

\bibliography{apssamp}

\providecommand{\noopsort}[1]{}\providecommand{\singleletter}[1]{#1}%
\begin{thebibliography}{35}%
\makeatletter
\providecommand \@ifxundefined [1]{%
 \@ifx{#1\undefined}
}%
\providecommand \@ifnum [1]{%
 \ifnum #1\expandafter \@firstoftwo
 \else \expandafter \@secondoftwo
 \fi
}%
\providecommand \@ifx [1]{%
 \ifx #1\expandafter \@firstoftwo
 \else \expandafter \@secondoftwo
 \fi
}%
\providecommand \natexlab [1]{#1}%
\providecommand \enquote  [1]{``#1''}%
\providecommand \bibnamefont  [1]{#1}%
\providecommand \bibfnamefont [1]{#1}%
\providecommand \citenamefont [1]{#1}%
\providecommand \href@noop [0]{\@secondoftwo}%
\providecommand \href [0]{\begingroup \@sanitize@url \@href}%
\providecommand \@href[1]{\@@startlink{#1}\@@href}%
\providecommand \@@href[1]{\endgroup#1\@@endlink}%
\providecommand \@sanitize@url [0]{\catcode `\\12\catcode `\$12\catcode `\&12\catcode `\#12\catcode `\^12\catcode `\_12\catcode `\%12\relax}%
\providecommand \@@startlink[1]{}%
\providecommand \@@endlink[0]{}%
\providecommand \url  [0]{\begingroup\@sanitize@url \@url }%
\providecommand \@url [1]{\endgroup\@href {#1}{\urlprefix }}%
\providecommand \urlprefix  [0]{URL }%
\providecommand \Eprint [0]{\href }%
\providecommand \doibase [0]{https://doi.org/}%
\providecommand \selectlanguage [0]{\@gobble}%
\providecommand \bibinfo  [0]{\@secondoftwo}%
\providecommand \bibfield  [0]{\@secondoftwo}%
\providecommand \translation [1]{[#1]}%
\providecommand \BibitemOpen [0]{}%
\providecommand \bibitemStop [0]{}%
\providecommand \bibitemNoStop [0]{.\EOS\space}%
\providecommand \EOS [0]{\spacefactor3000\relax}%
\providecommand \BibitemShut  [1]{\csname bibitem#1\endcsname}%
\let\auto@bib@innerbib\@empty
\bibitem [{\citenamefont {{IPPOG Collaboration}}()}]{AuthorList}%
  \BibitemOpen
  \bibfield  {author} {\bibinfo {author} {\bibnamefont {{IPPOG Collaboration}}},\ }\href@noop {} {}\bibinfo {howpublished} {\url{https://cds.cern.ch/record/2925772/files/IPPOG-AuthorList-may2025.pdf}}\BibitemShut {NoStop}%
\bibitem [{\citenamefont {{IPPOG International Masterclasses}}()}]{ippog-imc}%
  \BibitemOpen
  \bibfield  {author} {\bibinfo {author} {\bibnamefont {{IPPOG International Masterclasses}}},\ }\href@noop {} {}\bibinfo {howpublished} {\url{https://ippog.org/imc-international-masterclasses}},\ \bibinfo {note} {[accessed 2025-07-21]}\BibitemShut {NoStop}%
\bibitem [{\citenamefont {{IPPOG Global Cosmics Activities}}()}]{GlobalCosmics}%
  \BibitemOpen
  \bibfield  {author} {\bibinfo {author} {\bibnamefont {{IPPOG Global Cosmics Activities}}},\ }\href@noop {} {}\bibinfo {howpublished} {\url{https://ippog.org/global-cosmic-rays-portal}},\ \bibinfo {note} {accessed: 2025-07-21}\BibitemShut {NoStop}%
\bibitem [{\citenamefont {{HEP Outreach Projects Portal}}()}]{ResourcesPortal}%
  \BibitemOpen
  \bibfield  {author} {\bibinfo {author} {\bibnamefont {{HEP Outreach Projects Portal}}},\ }\href@noop {} {}\bibinfo {howpublished} {\url{https://ippog-resources-portal.web.cern.ch/}},\ \bibinfo {note} {accessed: 2025-07-21}\BibitemShut {NoStop}%
\bibitem [{\citenamefont {{Michael Kobel, TU Dresden}}(2005)}]{ippog-imc2005}%
  \BibitemOpen
  \bibfield  {author} {\bibinfo {author} {\bibnamefont {{Michael Kobel, TU Dresden}}},\ }\href@noop {} {\bibinfo {title} {{IMC Spreads the Word for Physics}}},\ \bibinfo {howpublished} {\url{https://cerncourier.com/a/masterclass-spreads-the-word-for-physics/}} (\bibinfo {year} {2005})\BibitemShut {NoStop}%
\bibitem [{\citenamefont {{QuarkNet}}()}]{QuarkNet}%
  \BibitemOpen
  \bibfield  {author} {\bibinfo {author} {\bibnamefont {{QuarkNet}}},\ }\href@noop {} {\bibinfo {title} {{Cosmic Rays Activities}}},\ \bibinfo {howpublished} {\url{https://quarknet.org/}},\ \bibinfo {note} {accessed: 2025-07-21}\BibitemShut {NoStop}%
\bibitem [{\citenamefont {{IPPOG}}(2024{\natexlab{a}})}]{IPPOGgeneralreport}%
  \BibitemOpen
  \bibfield  {author} {\bibinfo {author} {\bibnamefont {{IPPOG}}},\ }\href@noop {} {\bibinfo {title} {{General Report}}},\ \bibinfo {howpublished} {\url{https://cds.cern.ch/record/2930964/files/General Report.pdf}} (\bibinfo {year} {2024}{\natexlab{a}})\BibitemShut {NoStop}%
\bibitem [{\citenamefont {{IPPOG}}(2024{\natexlab{b}})}]{IPPOG-membersreport}%
  \BibitemOpen
  \bibfield  {author} {\bibinfo {author} {\bibnamefont {{IPPOG}}},\ }\href@noop {} {\bibinfo {title} {{Members Report}}},\ \bibinfo {howpublished} {\url{https://cds.cern.ch/record/2930964/files/IPPOG Members Report.pdf}} (\bibinfo {year} {2024}{\natexlab{b}})\BibitemShut {NoStop}%
\bibitem [{\citenamefont {Johansson}\ \emph {et~al.}(2006)\citenamefont {Johansson} \emph {et~al.}}]{Johansson:2006}%
  \BibitemOpen
  \bibfield  {author} {\bibinfo {author} {\bibfnamefont {K.}~\bibnamefont {Johansson}} \emph {et~al.},\ }\bibfield  {title} {\bibinfo {title} {{Hands on CERN: a well-used physics education project}},\ }\href {https://doi.org/10.1088/0031-9120/41/3/007} {\bibfield  {journal} {\bibinfo  {journal} {Phys. Educ.}\ }\textbf {\bibinfo {volume} {41}},\ \bibinfo {pages} {250} (\bibinfo {year} {2006})}\BibitemShut {NoStop}%
\bibitem [{\citenamefont {Kobel}\ \emph {et~al.}(2007)\citenamefont {Kobel} \emph {et~al.}}]{Kobel:2007}%
  \BibitemOpen
  \bibfield  {author} {\bibinfo {author} {\bibfnamefont {M.}~\bibnamefont {Kobel}} \emph {et~al.},\ }\bibfield  {title} {\bibinfo {title} {{European particle physics masterclasses make students into scientists for a day}},\ }\href {https://doi.org/10.1088/0031-9120/42/6/012} {\bibfield  {journal} {\bibinfo  {journal} {Phys. Educ.}\ }\textbf {\bibinfo {volume} {42}},\ \bibinfo {pages} {636} (\bibinfo {year} {2007})}\BibitemShut {NoStop}%
\bibitem [{\citenamefont {{M. Kobel \textit{et al.}}}(2014)}]{LEP}%
  \BibitemOpen
  \bibfield  {author} {\bibinfo {author} {\bibnamefont {{M. Kobel \textit{et al.}}}},\ }\href@noop {} {\bibinfo {title} {{How the Particle Physics Masterclasses began}}},\ \bibinfo {howpublished} {\url{https://cerncourier.com/a/how-the-particle-physics-masterclasses-began/}} (\bibinfo {year} {2014})\BibitemShut {NoStop}%
\bibitem [{\citenamefont {{IPPOG IMC Steering Group}}(2014)}]{ippog-imc2014}%
  \BibitemOpen
  \bibfield  {author} {\bibinfo {author} {\bibnamefont {{IPPOG IMC Steering Group}}},\ }\href@noop {} {\bibinfo {title} {{Masterclasses in the LHC era}}},\ \bibinfo {howpublished} {\url{https://cerncourier.com/a/international-masterclasses-in-the-lhc-era/}} (\bibinfo {year} {2014})\BibitemShut {NoStop}%
\bibitem [{\citenamefont {{ATLAS Masterclasses}}()}]{ATLASMasterclasses}%
  \BibitemOpen
  \bibfield  {author} {\bibinfo {author} {\bibnamefont {{ATLAS Masterclasses}}},\ }\href@noop {} {}\bibinfo {howpublished} {\url{https://atlas.cern/Resources/Particle-Physics-Masterclasses}},\ \bibinfo {note} {accessed: 2025-07-15}\BibitemShut {NoStop}%
\bibitem [{\citenamefont {{CMS Masterclasses}}()}]{CMSMasterclasses}%
  \BibitemOpen
  \bibfield  {author} {\bibinfo {author} {\bibnamefont {{CMS Masterclasses}}},\ }\href@noop {} {}\bibinfo {howpublished} {\url{https://cms-masterclass.web.cern.ch/}},\ \bibinfo {note} {accessed: 2025-07-15}\BibitemShut {NoStop}%
\bibitem [{\citenamefont {{ALICE Masterclasses}}()}]{ALICEMasterclasses}%
  \BibitemOpen
  \bibfield  {author} {\bibinfo {author} {\bibnamefont {{ALICE Masterclasses}}},\ }\href@noop {} {}\bibinfo {howpublished} {\url{https://alice-web-masterclass.app.cern.ch/home}},\ \bibinfo {note} {accessed: 2025-07-15}\BibitemShut {NoStop}%
\bibitem [{\citenamefont {{LHCb Masterclasses}}()}]{LHCbD0LifetimeMasterclass}%
  \BibitemOpen
  \bibfield  {author} {\bibinfo {author} {\bibnamefont {{LHCb Masterclasses}}},\ }\href@noop {} {}\bibinfo {howpublished} {\url{https://lhcb-outreach.web.cern.ch/lhcbinternationalmasterclasses/d0-lifetime/}},\ \bibinfo {note} {accessed: 2025-07-15}\BibitemShut {NoStop}%
\bibitem [{\citenamefont {{Belle II Masterclasses}}()}]{BELLEIIMasterclasses}%
  \BibitemOpen
  \bibfield  {author} {\bibinfo {author} {\bibnamefont {{Belle II Masterclasses}}},\ }\href@noop {} {}\bibinfo {howpublished} {\url{https://belle2.ijs.si/public/}},\ \bibinfo {note} {accessed: 2025-07-15}\BibitemShut {NoStop}%
\bibitem [{\citenamefont {{MINER$\,\nu$A Neutrino Masterclass}}()}]{MINERVAMasterclass}%
  \BibitemOpen
  \bibfield  {author} {\bibinfo {author} {\bibnamefont {{MINER$\,\nu$A Neutrino Masterclass}}},\ }\href@noop {} {}\bibinfo {howpublished} {\url{https://indico.fnal.gov/event/22340}},\ \bibinfo {note} {accessed: 2025-07-15}\BibitemShut {NoStop}%
\bibitem [{\citenamefont {{NO$\nu$A Neutrino Masterclass}}(2024)}]{NOVAMasterclass2024}%
  \BibitemOpen
  \bibfield  {author} {\bibinfo {author} {\bibnamefont {{NO$\nu$A Neutrino Masterclass}}},\ }\href@noop {} {}\bibinfo {howpublished} {\url{https://indico.fnal.gov/event/63011/}} (\bibinfo {year} {2024}),\ \bibinfo {note} {event held March 1--27, 2024; Accessed: 2025-07-15}\BibitemShut {NoStop}%
\bibitem [{\citenamefont {{Pierre Auger Collaboration Masterclasses}}(2023)}]{AugerMasterclasses}%
  \BibitemOpen
  \bibfield  {author} {\bibinfo {author} {\bibnamefont {{Pierre Auger Collaboration Masterclasses}}},\ }\href@noop {} {}\bibinfo {howpublished} {\url{https://pos.sissa.it/444/1611/}} (\bibinfo {year} {2023}),\ \bibinfo {note} {38th International Cosmic Ray Conference (ICRC2023)}\BibitemShut {NoStop}%
\bibitem [{\citenamefont {{Pierre Auger Collaboration Masterclasses}}()}]{AugerMasterclassesWeb}%
  \BibitemOpen
  \bibfield  {author} {\bibinfo {author} {\bibnamefont {{Pierre Auger Collaboration Masterclasses}}},\ }\href@noop {} {}\bibinfo {howpublished} {\url{https://augermasterclasses.lip.pt/}},\ \bibinfo {note} {accessed: 2025-07-15}\BibitemShut {NoStop}%
\bibitem [{\citenamefont {{Panagiota Foka, GSI}}()}]{ParticleTherapyMasterClass2020}%
  \BibitemOpen
  \bibfield  {author} {\bibinfo {author} {\bibnamefont {{Panagiota Foka, GSI}}},\ }\href@noop {} {\bibinfo {title} {{Particle Therapy MasterClass}}},\ \bibinfo {howpublished} {\url{https://pos.sissa.it/398/910/pdf}},\ \bibinfo {note} {2021 EPS-HEP Conference}\BibitemShut {NoStop}%
\bibitem [{\citenamefont {{Despina Hatzifotiadou}}()}]{EngagingWorld}%
  \BibitemOpen
  \bibfield  {author} {\bibinfo {author} {\bibnamefont {{Despina Hatzifotiadou}}},\ }\href@noop {} {\bibinfo {title} {{Engaging the world with science}}},\ \bibinfo {howpublished} {\url{https://pos.sissa.it/466/014/pdf}},\ \bibinfo {note} {2024 Latice Conference}\BibitemShut {NoStop}%
\bibitem [{\citenamefont {Tejeda~Mu{\~n}oz}\ \emph {et~al.}(2025)\citenamefont {Tejeda~Mu{\~n}oz}, \citenamefont {Perez~Moreno}, \citenamefont {Ragoni}, \citenamefont {Regules~Medel}, \citenamefont {Fern\'andez~T\'ellez},\ and\ \citenamefont {Vasquez~Beltran}}]{TejedaMunoz2025CosmicPiano}%
  \BibitemOpen
  \bibfield  {author} {\bibinfo {author} {\bibfnamefont {G.}~\bibnamefont {Tejeda~Mu{\~n}oz}}, \bibinfo {author} {\bibfnamefont {L.~A.}\ \bibnamefont {Perez~Moreno}}, \bibinfo {author} {\bibfnamefont {S.}~\bibnamefont {Ragoni}}, \bibinfo {author} {\bibfnamefont {H.~D.}\ \bibnamefont {Regules~Medel}}, \bibinfo {author} {\bibfnamefont {A.}~\bibnamefont {Fern\'andez~T\'ellez}},\ and\ \bibinfo {author} {\bibfnamefont {Y.~A.}\ \bibnamefont {Vasquez~Beltran}},\ }\bibfield  {title} {\bibinfo {title} {Cosmic piano: a modular scintillator-based muon detector for scientific outreach},\ }\href {https://doi.org/10.1088/1361-6552/adc2c5} {\bibfield  {journal} {\bibinfo  {journal} {Physics Education}\ }\textbf {\bibinfo {volume} {60}},\ \bibinfo {pages} {035031} (\bibinfo {year} {2025})}\BibitemShut {NoStop}%
\bibitem [{\citenamefont {{Carolin Schwerdt, DESY}}()}]{CosMO}%
  \BibitemOpen
  \bibfield  {author} {\bibinfo {author} {\bibnamefont {{Carolin Schwerdt, DESY}}},\ }\href@noop {} {\bibinfo {title} {{CosMO – A Cosmic Muon Observer Experiment for Students}}},\ \bibinfo {howpublished} {\url{https://arxiv.org/pdf/1309.3391}},\ \bibinfo {note} {2013 ICRC Conference}\BibitemShut {NoStop}%
\bibitem [{\citenamefont {{Philipp Lindenau, Carolin Schwerdt and Michael Walter}}()}]{Cosmic@Web}%
  \BibitemOpen
  \bibfield  {author} {\bibinfo {author} {\bibnamefont {{Philipp Lindenau, Carolin Schwerdt and Michael Walter}}},\ }\href@noop {} {\bibinfo {title} {{Students work like astroparticle physicists with Cosmic@Web}}},\ \bibinfo {howpublished} {\url{https://pos.sissa.it/395/1398/pdf}},\ \bibinfo {note} {2021 ICRC Conference}\BibitemShut {NoStop}%
\bibitem [{\citenamefont {{Accel Kitchen LLC}}()}]{Accel}%
  \BibitemOpen
  \bibfield  {author} {\bibinfo {author} {\bibnamefont {{Accel Kitchen LLC}}},\ }\href@noop {} {\bibinfo {title} {{Online support for research activities by high school and junior high school students}}},\ \bibinfo {howpublished} {\url{https://pos.sissa.it/444/1600/pdf}},\ \bibinfo {note} {2023 ICRC Conference}\BibitemShut {NoStop}%
\bibitem [{\citenamefont {{Nicolas Arnaud et al}}()}]{cosmixcase}%
  \BibitemOpen
  \bibfield  {author} {\bibinfo {author} {\bibnamefont {{Nicolas Arnaud et al}}},\ }\href@noop {} {\bibinfo {title} {{Cosmos à l’École}}},\ \bibinfo {howpublished} {\url{https://indico.cern.ch/event/1258933/contributions/6475923/attachments/3105910/5504550/20250718_ICRC.pdf}},\ \bibinfo {note} {2025 International Cosmic Ray Conference}\BibitemShut {NoStop}%
\bibitem [{\citenamefont {{QuarkNet}}(nd)}]{MuonWeek}%
  \BibitemOpen
  \bibfield  {author} {\bibinfo {author} {\bibnamefont {{QuarkNet}}},\ }\href@noop {} {\bibinfo {title} {{International Muon Week}}},\ \bibinfo {howpublished} {\url{https://quarknet.org/content/international-muon-week}} (\bibinfo {year} {n.d.}),\ \bibinfo {note} {accessed: 2025-07-21}\BibitemShut {NoStop}%
\bibitem [{\citenamefont {Aramo}\ and\ \citenamefont {Hemmer}(2020)}]{Aramo:2020les}%
  \BibitemOpen
  \bibfield  {author} {\bibinfo {author} {\bibfnamefont {C.}~\bibnamefont {Aramo}}\ and\ \bibinfo {author} {\bibfnamefont {S.}~\bibnamefont {Hemmer}} (\bibinfo {collaboration} {OCRA}),\ }\bibfield  {title} {\bibinfo {title} {{Outreach Cosmic Ray Activities (OCRA): a program of Astroparticle Physics Outreach Events for High-School Students}},\ }\href {https://doi.org/10.22323/1.358.0173} {\bibfield  {journal} {\bibinfo  {journal} {PoS}\ }\textbf {\bibinfo {volume} {ICRC2019}},\ \bibinfo {pages} {173} (\bibinfo {year} {2020})}\BibitemShut {NoStop}%
\bibitem [{\citenamefont {Pisano}(2022)}]{Pisano:2022}%
  \BibitemOpen
  \bibfield  {author} {\bibinfo {author} {\bibfnamefont {S.}~\bibnamefont {Pisano}} (\bibinfo {collaboration} {EEE}),\ }\bibfield  {title} {\bibinfo {title} {{The extreme energy events project. Bring science inside schools}},\ }\href {https://doi.org/10.1140/epjp/s13360-022-03331-0} {\bibfield  {journal} {\bibinfo  {journal} {Eur. Phys. J. Plus}\ }\textbf {\bibinfo {volume} {137}},\ \bibinfo {pages} {1190} (\bibinfo {year} {2022})}\BibitemShut {NoStop}%
\bibitem [{\citenamefont {{International Cosmic Day}}(2025)}]{ICD}%
  \BibitemOpen
  \bibfield  {author} {\bibinfo {author} {\bibnamefont {{International Cosmic Day}}},\ }\href@noop {} {}\bibinfo {howpublished} {\url{https://icd.desy.de/}} (\bibinfo {year} {2025}),\ \bibinfo {note} {accessed: 2025-07-21}\BibitemShut {NoStop}%
\bibitem [{\citenamefont {{IPPOG RDB}}(2025)}]{RDB}%
  \BibitemOpen
  \bibfield  {author} {\bibinfo {author} {\bibnamefont {{IPPOG RDB}}},\ }\href@noop {} {}\bibinfo {howpublished} {\url{https://ippog.org/ippog-resource-database}} (\bibinfo {year} {2025}),\ \bibinfo {note} {accessed: 2025-07-21}\BibitemShut {NoStop}%
\bibitem [{\citenamefont {{IPPOG Partners web sites}}()}]{WebSites}%
  \BibitemOpen
  \bibfield  {author} {\bibinfo {author} {\bibnamefont {{IPPOG Partners web sites}}},\ }\href@noop {} {}\bibinfo {howpublished} {\url{https://ippog.org/resource-websites}},\ \bibinfo {note} {accessed: 2025-07-21}\BibitemShut {NoStop}%
\bibitem [{\citenamefont {{CERN}}(2025)}]{TeachersProgramme}%
  \BibitemOpen
  \bibfield  {author} {\bibinfo {author} {\bibnamefont {{CERN}}},\ }\href@noop {} {\bibinfo {title} {{CERN Teachers Programme}}},\ \bibinfo {howpublished} {\url{https://teachers.cern/}} (\bibinfo {year} {2025}),\ \bibinfo {note} {accessed: 2025-07-21}\BibitemShut {NoStop}%
\end{thebibliography}%

\clearpage
\appendix
\section*{Appendix A: IPPOG Collaboration Author List}

\noindent
Pedro Abreu$^{1}$, Claire Adam$^{2}$, Calin Alexa$^{3}$, Muhammad Alhroob$^{4}$, Boyka Aneva$^{5}$, Nicolas Angelides$^{6}$, Iqbal Muhammad Ansar$^{7}$, Katarina Anthony$^{8}$, Nicolas Arnaud$^{9}$, Ralf Averbeck$^{10}$
% ...

\vspace{1.2em}
\noindent\textbf{Affiliations}\\
$^{1}$ LIP — Laboratorio de Instrumentação e Física Experimental de Partículas; and IST — University of Lisboa, Portugal \\
$^{2}$ CNRS/IN2P3, LAPP - Annecy, France \\
$^{3}$ IFIN-HH, Bucharest, Romania \\
$^{4}$ University of Warwick, United Kingdom \\
$^{5}$ Bulgarian Academy of Science, Bulgaria \\
$^{6}$ University of Zurich, Switzerland \\
$^{7}$ University of California, Los Angeles, USA \\
$^{8}$ CERN, Geneva, Switzerland \\
$^{9}$ CNRS/IN2P3, IP2I, Université Claude Bernard Lyon 1, France \\
$^{10}$ GSI — Helmholtzzentrum für Schwerionenforschung GmbH, Germany \\
% ...
$^{46}$ CMS Collaboration, CERN (Geneva, Switzerland) \\
% ...

\subsection*{Financial Support}

The IPPOG coordinators are supported by the following institutes: CERN, CNRS/IN2P3 (F), DESY (DE), INFN (IT), LIP (PT), University of Notre Dame (US).
\vspace{1em}

The IPPOG collaboration is financially supported by the following entities: 

\begin{itemize}
  \item Australia: CoEPP, University of Melbourne
  \item Austria: Oesterreichiche Physikalische Gesellschaft / Stefan Meyer Institute / Institute of High Energy Physics, Austrian Academy of Sciences
  \item Belgium: UCLouvain
  \item Brazil: Renafae / CBPF
  \item Bulgaria: Sofia Tech Park Jsc
  \item CERN: European Organization for Nuclear Research
  \item Cyprus: University of Cyprus
  \item Czech Republic: Institute of Physics of the Czech Academy of Sciences
  \item Denmark: Niels Bohr Institute
  \item Finland: Helsinki Institute of Physics (HIP)
  \item France: Centre National de la Recherche Scientifique (CNRS)
  \item Georgia: Ministry of Education and Science of Georgia
  \item Germany: Deutsches Elektronen-Synchrotron (DESY)
  \item Greece: Ministry of Development and Investment
  \item GSI: GSI Helmholtzzentrum für Schwerionenforschung
  \item Hungary: Wigner Research Centre for Physics
  \item India: National Institute of Science, Education and Research (NISER)
  \item Ireland: Dublin Institute for Advanced Studies
  \item Israel: Weizmann Institute of Science / Tel Aviv University
  \item Italy: INFN
  \item Latvia: Riga Technical University
  \item Mexico: Benemérita Universidad Autónoma de Puebla
  \item Montenegro: Ministry of Education, Science and Innovation
  \item Netherlands: Nikhef – National Institute
  \item Norway: ATLAS Norway
  \item Poland: University of Warsaw
  \item Portugal: LIP – Laboratorio de Instrumentação e Física Experimental de Partículas
  \item Romania: IFA
  \item Slovakia: Ministry of Education, Research, Development and Youth
  \item Slovenia: University of Ljubljana
  \item South Africa: NRF – iThemba
  \item Spain: CIEMAT, Madrid
  \item Sweden: KTH, Stockholm
  \item Switzerland: Physik-Institut der Universität Zürich
  \item UK: Science and Technology Facilities Council
  \item USA: QuarkNet Programme, University of Notre Dame
\end{itemize}

\end{document}